\documentclass[twocolumn,aps,prb,showpacs]{revtex4-1}

\usepackage{amsmath}
\usepackage{amssymb}
\usepackage{upgreek}
\usepackage[tight,TABTOPCAP]{subfigure}
\usepackage{float}
\usepackage{graphicx}
\usepackage{wrapfig}
\usepackage{natbib}
\usepackage{color}

\newcommand{\vect}[1]{{\boldsymbol{#1}}}

\makeatletter

\begin{document}

\title{Effect of Dynamical Screening on Single Particle Spectral Features of Uniaxially Strained Graphene: Tuning the Plasmaron Ring}

\author{J. P. F. LeBlanc$^{1}$}
\email{jpfleblanc@gmail.com}
\author{J. P. Carbotte$^{2,3}$}

\affiliation{$^1$Max-Planck-Institute for the Physics of Complex Systems, 01187 Dresden, Germany} 
\affiliation{$^2$Department of Physics and Astronomy, McMaster
University, Hamilton, Ontario L8S~4M1 Canada}
\affiliation{$^3$The Canadian Institute for Advanced Research, Toronto, ON M5G~1Z8 Canada}


\date{\today}
\begin{abstract}
Electronic screening renormalizes the linear bands of graphene and in the vicinity of the Dirac point, creates a diamond shaped structure in the quasiparticle spectral density.  This is a result of electron-plasmon scattering processes which produce a finite momentum feature referred to as the `plasmaron ring'.  In this work we explore the effects of uniaxial strain on these spectral features with the aim of understanding how strain modifies correlations.  We derive and calculate the spectral density to the G$_0$W-RPA level which allows us to identify the dispersive behaviour of the diamond geometry, and thus electron-plasmon scattering, for variation in electron-electron coupling strength and magnitude of applied strain.
We find that the application of strain changes the geometry (in momentum) of the electron-plasmon scattering and that renormalizations beyond simple geometrical scalings further enhance this effect.  These results suggest that the properties of the plasmaron ring can be tuned through the application of uniaxial strain, effectively producing a larger fine structure constant without the need to change the sample substrate.  
\end{abstract}

\maketitle

\section{Introduction}
Graphene is a two-dimensional system comprised of a single layer of carbon in a honeycomb lattice.  Near specific points in reciprocal space, named $K$ and $K^\prime$ valleys, the dispersion is conical, terminating at a Dirac point in each valley.
Angle-resolved photoemission spectroscopy (ARPES) work by Bostwick et al.\cite{bostwick:2010} has resulted in a significant interest in many-body interactions and their effects near the Dirac point of graphene.  In that work, and others,\cite{walter:2011} it was observed that, for certain substrates, the single Dirac point splits into two pieces.  In addition to this, the dispersion away from the K and K$^\prime$ points  shows new structure due to electronic scattering with plasmons.  This additional energy structure which occurs between the two Dirac-like points, illustrated schematically in Fig.~\ref{fig:1}(a) at energy $E_1$, has been referred to in the literature as a plasmaron ring.\cite{bostwick:2010,walter:2011}  This structure has been shown to be well described by the G$_0$W-RPA calculation of the screened electron-electron interaction.
Further, there is some speculation that these additional energy structures, whose physical origin are a strong electron-plasmon scattering resonance, may be useful in the design of plasmonic devices.\cite{grigorenko:2012}

Recent infrared near-field optical techniques\cite{fei:2012, chen:2012,grigorenko:2012, carbotte:2012} have managed to observe plasmon dispersions and are hoped to be complementary to existing ARPES work.  Since these spectroscopic techniques are in general a momentum averaged probe, it is essential to have multiple tuning mechanisms in order to identify and separate the impacts of different many-body effects on spectral features as has been done in ARPES.\cite{bostwick:2007,leblanc:2012, jungseek:2012}  The most commonly exploited tuning parameter is the chemical potential which can be controlled through gating as done, for example, in optics.\cite{li:Basov:2009, fei:2012}  In the case of electron-electron interactions, the substrate also plays a substantial role, as it modifies the effective fine structure constant, $\alpha=\frac{g e^2}{\upvarepsilon_0 v_F}$, through a change in the value of the effective dielectric constant of graphene at the substrate boundary, $\upvarepsilon_0$.  Here $g$ is the spin-valley degeneracy factor ($g=g_s g_v=4$), $e$ is the electron charge and $v_F$ is the Fermi velocity.  Since $\alpha$ is dependent upon the graphene substrate, tuning $\alpha$ is difficult in experiment, as one is somewhat limited in the values of $\alpha$ one can obtain from otherwise suitable substrates.\cite{walter:2011} 

There is recent interest in strain on Dirac fermions in artificial honeycomb lattices produced in cold atom systems, which in some experimental set-ups naturally produce a strained Dirac dispersion.\cite{tarruell:2012, uehlinger:2012}  In physical graphene, deformations as large as 20\% \cite{kim:2009, liu:2007, cadelano:2009} are possible above which an energy gap opens in the electronic spectrum.  It may also be the case that strain can act as a tuning parameter for certain measurable quantities and could be an important experimental tool.
In this paper we derive and calculate the G$_0$W-RPA result for the screened electron-electron interaction (EEI) in doped graphene for arbitrary magnitude and direction of strain.  We find that the application of strain modulates the radius of the plasmaron ring in $k$-space and that the electron-electron interaction further enhances this effect.  These results suggest that the properties of the plasmaron ring can be tuned and therefore probed through the application of uniaxial strain, producing an effectively larger fine structure constant along the strained axis without the need to change the sample substrate.  
We begin with the basic formulas for a strained Dirac Hamiltonian in Sec.~\ref{sec:strain}.  In Sec.~\ref{sec:selfe} we discuss how the strain affects existing theory on the G$_0$-RPA level, Sec.~\ref{sec:results} will contain our main results, and Sec.~\ref{sec:conclusions} will conclude.

\section{Strain}\label{sec:strain}

The low energy Hamiltonian for a strained graphene sheet can be written in a linear form for an effective momentum, $\vect{\bar{k}}$, as
\begin{equation}
H=\hbar v_{F}\boldsymbol{\sigma \cdot \bar{k}}.
\end{equation}
From this starting point, a system with strain applied at an arbitrary angle $\gamma$ can be related to a symmetric system through the transformation\cite{pellegrino:2011} 
\begin{equation}
\boldsymbol{\bar{k}}= A(\gamma)\boldsymbol{k},
\end{equation}
where $A(\gamma)$ is the product of rotations and strain given by
\begin{equation}\label{eqn:strain}
A(\gamma) = R(\gamma)S(\upvarepsilon)R(-\gamma).
\end{equation}
$S(\epsilon)$ is a 2x2 straining matrix and $R(\gamma)$ is a standard 2-dimensional rotation through the angle $\gamma$.
In the case of uniaxial strain, the strain matrix is a scaling in $k_x$ and $k_y$ given simply as
\begin{equation}
S(\upvarepsilon)=\begin{pmatrix}
c_\parallel & 0 \\
0 & c_\perp
\end{pmatrix}
\end{equation}
where $c_\parallel$ and $c_\perp$ are interrelated through some material specific constants: $\kappa=\frac{a}{2 t}(\frac{\partial t}{\partial a})-\frac{1}{2}$, related to the change to nearest-neighbour hopping, $t$, with a modification of the lattice spacing, $a$; Poisson's ratio $\nu$, the negative ratio of transverse to axial strain; and the strain modulus, $\epsilon$. We have taken these to have values of $\kappa=1.1$ and $\nu=0.14$ according to Ref.~\onlinecite{farjam:2009} and arbitrary strain modulus.  The relation to strain factors is given by
\begin{align}
c_\parallel &= 1-2\kappa \epsilon , \\
c_\perp &= 1+2\kappa \nu \epsilon .
\end{align}
\begin{figure}
  \begin{center}
  \includegraphics[width=0.95\linewidth]{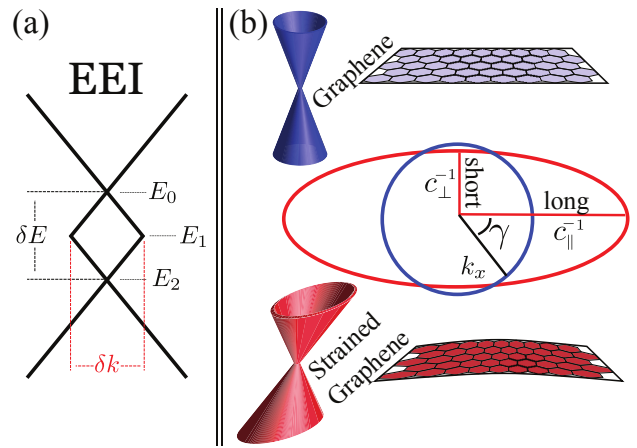}
  \end{center}
  \caption{\label{fig:1}(Color online) (a)  Inclusion of electron-electron interactions creates a diamond structure with energy width $\delta E$ and momentum radius, $\delta k$, which describes the plasmaron ring. (b)Schematic illustrating a conical Dirac dispersion (blue) strained along an arbitrary angle $\gamma$ (red), with long and short axis factors $c_\parallel$ and $c_\perp$. }
\end{figure}
We illustrate schematically in Fig.~\ref{fig:1}(b) how a conical dispersion, with a circular cross section, when strained at an arbitrary angle $\gamma$ from the $k_x$-direction produces an elliptical cross section.
For any direction of strain, $A(\gamma)$ is a 2x2 matrix, with elements $A_{ij}$ defined by the matrix multiplications of Eq.~(\ref{eqn:strain}), which modifies the momenta $\vect{k}$.
  One can then write an arbitrary vector $\boldsymbol{\bar{k}}$ as
\begin{equation}\label{eqn:B}
\boldsymbol{\bar{k}}= k\begin{pmatrix}
A_{11}\cos\theta_{\boldsymbol{k}}+A_{12}\sin\theta_{\boldsymbol{k}} \\ A_{21}\cos\theta_{\boldsymbol{k}}+A_{22}\sin\theta_{\boldsymbol{k}}
\end{pmatrix} = k\begin{pmatrix}
B_x(\theta_{\vect{k}}) \\ B_y(\theta_{\vect{k}})
\end{pmatrix}.
\end{equation}
Here we have introduced a short hand for the $x$ and $y$ axis modifications due to strain, such that $\boldsymbol{\bar{k}}= k \vect{B}$.
For a given $\epsilon$ and $\gamma$, the $A_{ij}$ are constant and one can then write a vector in terms of its strain contracted magnitude and direction.  This allows for easy replacement of both the length and the angle of any unstrained vector with its contracted values.
While in principle $\vect{B}$ contains strain modulus, strain angle and momentum angle dependence, for a given system we assume that $\epsilon$ and $\gamma$ are fixed, and so will refer to $\vect{B}$ as having only angular dependence.  The length contraction is therefore simply
\begin{equation}
|\vect{B}(\theta_{\boldsymbol{k}})|=\sqrt{B_x^2(\theta_{\vect{k}}) +B_y^2(\theta_{\vect{k}})}
\end{equation}
where $B_{x}$ and $B_{y}$ are defined by Eq.~(\ref{eqn:B}) for a given angle $\theta_{\vect{k}}$.
The resulting angle of the strained vector $\vect{\bar{k}}$ is therefore
\begin{equation}\label{eqn:mag}
\beta_{\overline{\vect{k}}}= \arctan\left( \frac{B_y(\theta_{\vect{k}})}{B_x(\theta_{\vect{k}})}\right).
\end{equation}
Note that for improved clarity, we use the notation that barred variables have angles of $\beta$ rather than $\theta$.
Simple manipulation would show that in the unstrained case, where $c_\parallel=c_\perp=1$, you get the original magnitude and angle.  Further, the momentum length contraction and angle shift are dependent upon the angle of the vector and of the applied strain.
  While these formulas and those that follow in Section~\ref{sec:selfe} are written for arbitrary momentum and strain direction, it should be apparent that the cases of interest are for momentum either in the direction of the strain or transverse to the strain which we will refer to respectively as the long axis and the short axis.
  Depending on the application, one may wish to work purely in the principle axis coordinate system given by $\gamma=0$, which results in $A(\gamma=0)=S(\epsilon)$.  In this special reference frame, the strained momentum is
  \begin{equation}\label{eqn:straincoords}
  \boldsymbol{\bar{k}}= k\begin{pmatrix}
 c_\parallel \cos\theta_{\boldsymbol{k}} \\  c_\perp \sin\theta_{\boldsymbol{k}}
\end{pmatrix} = \begin{pmatrix}
c_\parallel k_x \\ c_\perp k_y
\end{pmatrix}.
  \end{equation}
  which is equivalent to the notation used elsewhere.\cite{pereira:2010}

\section{Dynamical Screening and Self Energies}\label{sec:selfe}
It has been shown in general\cite{pellegrino:2011} that the density-density correlation function in the strained case, $\Pi(\vect{q},\omega)$, is related to its unstrained counterpart, $\Pi^0(\vect{q},\omega)$, through
\begin{equation}\label{eqn:pifix}
\Pi(\boldsymbol{q},\omega)= [\det S(\epsilon)]^{-1} \Pi^0(\vect{\bar{q}},\omega).
\end{equation}
We see that the impact of strain is to both modulate the scattered momentum, $\boldsymbol{q}$, and also scale the polarization by the inverse determinant of the strain matrix, $S(\epsilon)$.
Here, in the case of uniaxial strain, the determinant is simply $c_\parallel c_\perp$, representative of the modified slope of the density of states at the Fermi level.  Further, since the polarization in the unstrained case only depends on the magnitude of $\boldsymbol{\overline{q}}$, the polarization in the strained case is evaluated at the scaled momentum given for a set of $\epsilon$ and $\gamma$ in Eq.~(\ref{eqn:B}).

The G$_0$W-RPA self energy for the screened electron-electron interaction for a given momentum, $\vect{k}$, and frequency, $\omega$, in unstrained graphene has been previously derived.\cite{polini:2008,leblanc:2011,hwang:2008}
The details of those derivations can be repeated, but with the added complexity of strain which results in
\begin{widetext}
\begin{align}\label{eqn:res}
{\Sigma}_s^{RES}(\vect{k},{\omega})=\sum_{s^\prime=\pm 1}\int\limits_{0}^\infty \int\limits_0^{2\pi} &\frac{d{q}d\theta_{\boldsymbol{q}}}{2\pi}\frac{\alpha}{g}\upvarepsilon^{-1}({q},{\omega}-{\epsilon}_{\vect{\overline{k+q}}}^{s^\prime})F_{ss^\prime}(\beta_{\boldsymbol{\bar{k}\bar{k}^\prime}}) \big[\Theta({\omega}-{\epsilon}_{\vect{\overline{k+q}}}^{s^\prime})-\Theta(-{\epsilon}_{\vect{\overline{k+q}}}^{s^\prime})\big].
\end{align}
and
\begin{align}\label{eqn:line}
{\Sigma}_s^{line}(\vect{k},{\omega})=&-\sum_{s^\prime=\pm 1}\int\limits_{0}^\infty \int\limits_{0}^{2\pi}\frac{d{q} d\theta_{\boldsymbol{q}}}{2\pi}\frac{\alpha}{g}F_{ss^\prime}(\beta_{\vect{\bar{k}\bar{k}}^\prime}) \int_{-\infty}^{\infty}\frac{d{\Omega}}{2\pi}\upvarepsilon^{-1}(q,i {\Omega})\Bigg[\frac{{\omega}-{\epsilon}_{\overline{\vect{k}+\vect{q}}}^{s^\prime}}{{\Omega}^2+({\epsilon}_{\overline{\vect{k}+\vect{q}}}^{s^\prime}-{\omega})^2}-\frac{i {\Omega}}{{\Omega}^2+({\epsilon}_{\overline{\vect{k}+\vect{q}}}^{s^\prime}-{\omega})^2}\Bigg].
\end{align}
\end{widetext}
This calculation primarily samples magnitudes of scattered $\vect{q}$ over all directions $d\theta_{\vect{q}}$.
The total self energy for a given band, $s=\pm 1$, is then the sum of these two contributions, $\Sigma_s(\vect{k},\omega)=\Sigma_s^{RES}(\vect{k},\omega)+\Sigma_s^{line}(\vect{k},\omega)$.  With the exception of the additional strained labels these formulas [Eqns.~(\ref{eqn:res}),(\ref{eqn:line})]   are identical to the unstrained self energies.\cite{polini:2008, leblanc:2011}  Nevertheless we will outline precisely how these strain factors influence the evaluation of functions contained within Eqs.~(\ref{eqn:res}-\ref{eqn:line}).
To begin, the prefactors are such that all quantities are scaled by the non-interacting chemical potential.  Thus, the strength of the EEI renormalization increases or decreases with the value of the chemical potential.  In the unstrained case, this scaling leads to momenta which are scaled by the Fermi momentum, $k_F$.  However, in the case of strain, since the slope of the dispersion is not uniform with angle, the Fermi momentum is instead given by the formula $\hbar k_F |\vect{B}(\theta_{\boldsymbol{k}})|=\mu$.  Thus, in the strained case, $k_F$ depends on the angle of $\vect{k}$ and is shifted from the unstrained case by a factor of $1/|\vect{B}(\theta_{\boldsymbol{k}})|$.  Also note that the upper limit of the $q$ integral requires a choice of cut-off to avoid an ultraviolet divergence.  In principal one might be concerned with a variation of this cut-off for different directions of $\vect{k}$ along the strained ellipse.  The choice of cut-off has been shown to enter renormalization properties logarithmically, and for large cut-off these variations will not make a significant contribution to the result.\cite{polini:2007}  Despite this, we make the simplest correction for this error by modulating the cut-off by an angularly dependent factor of $|B(\theta_{\vect{k}})|$.

We can see the intrinsic dependence in the self energy both on the direction of $\vect{k}$ and on the direction and strength of the applied strain.  These factors enter into the strained dispersion, ${\epsilon}_{\overline{\vect{k}+\vect{q}}}^{s^\prime}$, the inverse dielectric constant, $\upvarepsilon^{-1}(q,\omega)$, and the band overlap factors, $F_{ss^\prime}$.
The inverse dielectric constant is given by
\begin{equation}\label{eq:die}
\upvarepsilon^{-1}({q},{\omega})=\frac{1}{1-V_q\Pi(q,\omega)}=\frac{{q}}{{q}-\alpha {\Pi}(q, {\omega})},
\end{equation}
which through Eq.~(\ref{eqn:pifix}) becomes
\begin{equation}
\upvarepsilon^{-1}({q},{\omega})=\frac{{q}}{{q}-\alpha {\Pi^0}(\overline{q}, {\omega})/(c_\parallel c_\perp)},\label{eq:diefix}
\end{equation}
where $\Pi^0$ is the function in the unstrained case, but now evaluated at the strained momentum, $\bar{q}$, derived repeatedly elsewhere in explicit form.\cite{wunsch:2006,barlas:2007,leblanc:2011}
$V_q=\frac{2\pi e^2}{\upvarepsilon_0 q}$ is the two-dimensional coulomb potential for a given effective dielectric constant of the medium, $\upvarepsilon_0$.

The energies and wavefunctions are in the same form as in the unstrained case, but with strained variables for momentum and angle. The well known band overlap factor, $F_{ss^\prime}$, describing a transition from momentum $\vect{k}$ in the $s$ band to momentum $\vect{k^\prime}$ in the $s^\prime$ band acts to remove backscattering within a given band.  One obtains
\begin{equation}\label{eqn:12}
F_{ss^\prime}(\beta_{\boldsymbol{\bar{k}\bar{k^\prime}}})= \frac{1}{2}[1+ss^\prime \cos(\beta_{\boldsymbol{\bar{k}\bar{k^\prime}}})]
\end{equation}
where $\beta_{\boldsymbol{\bar{k}\bar{k}^\prime}}=\beta_{\boldsymbol{\bar{k}^\prime}}-\beta_{\boldsymbol{\bar{k}}}$ which  can immediately be written down from Eq.~(\ref{eqn:mag}).

For clarity, we show an example of how $\beta_{\boldsymbol{\bar{k}\bar{k}^\prime}}$ collapses for the special cases of a fixed vector along the long and short axes for the $\theta_{\vect{k}}=0$ case.  One requires two terms, the simplest is
\begin{equation}\label{eqn:17}
\beta_{\overline{\vect{k}}}= \arctan\left( \frac{B_y}{B_x}\right)=\arctan\left(\frac{A_{21}}{A_{11}}\right)=0
\end{equation}
since $A_{21}=0$ for $\gamma=0$ (long axis), and the same for $\gamma=\frac{\pi}{2}$ (short axis).  One could instead provide this example in the coordinates of the strain, as in Eq.~(\ref{eqn:straincoords}), which fixes $\gamma=0$.  In this case, one finds 
\begin{equation}\label{eqn:18}
\beta_{\overline{\vect{k}}}=\begin{cases} \arctan\left(\frac{A_{21}}{A_{11}}\right)=  0, & \text{long axis, $\theta_{\vect{k}}=0$} \\  \arctan\left(\frac{A_{22}}{A_{12}}\right)=  \frac{\pi}{2}, & \text{short axis, $\theta_{\vect{k}}=\frac{\pi}{2}$ }, \end{cases}
\end{equation}
a different result from Eq.~\ref{eqn:17}, but in both cases one obtains in Eqn.~(\ref{eqn:17}) and (\ref{eqn:18}) that $\beta_{\vect{\bar{k}}}=\theta_{\vect{k}}$, or that the strain transformation does not modify the vector angle for a momentum either in the long or short axial directions.

  The $\vect{\bar{k}^\prime}$ term is more complicated, as it depends on the angle to the vector $\vect{k^\prime}$.  Continuing our example for $\theta_{\vect{k}}=0$, one would obtain
\begin{equation}
\theta_{\vect{k^\prime}}=\arctan \left( \frac{q\sin\theta_{\vect{q}}}{k+ q\cos\theta_{\vect{q}}} \right)= \arctan \left( R \right)
\end{equation}
where we again have introduced a shorthand notation.  We can then write the desired angle
\begin{equation}
\beta_{\boldsymbol{\bar{k}^\prime}} = \arctan\left(\frac{A_{21}+A_{22}R}{A_{11}+A_{12}R}\right).
\end{equation}
Finally, in the case of no strain, $A_{21}=A_{12}=0$ and $A_{11}=A_{22}=1$ and we find that $\beta_{\vect{\bar{k}^\prime}}\to\theta_{\vect{k^\prime}}$.  In this limit the overlap function $F_{ss^\prime}(\beta_{\vect{\bar{k}\bar{k}^\prime}}) \to F_{ss^\prime}(\theta_{\vect{kk^\prime}})$ which is the standard result for unstrained graphene.\cite{wunsch:2006,neto:2009}
We can see from the above examples that one can align the coordinate system along either the direction of strain, $\gamma$, or along the momentum in question, $\vect{k}$.  All that matters for the calculation is the direction of $\vect{k}$ relative to the strain.  In practice, the two most interesting cases are for a momentum along the direction of strain (long axis) or perpendicular to it (short axis).

The energy dispersion for a band, $s$, is of the general form
\begin{equation}
\epsilon_{\boldsymbol{\overline{\vect{k}+\vect{q}}}}^s=s|\boldsymbol{\overline{\vect{k}+\vect{q}}}|-1
\end{equation}
which is in units of the bare chemical potential, $\mu$.
This can be written  out in terms of $\vect{k}$, $\vect{q}$ and the strain angle by using Eq.~(\ref{eqn:B}).
We can then write the strained Green's function for each band
\begin{equation}\label{eqn:gf}
G^s(\boldsymbol{k}, \omega)=\frac{1}{\omega-\epsilon_{\overline{\vect{k}}}^s - \Sigma_s(\boldsymbol{k},\omega)}.
\end{equation}
The total spectral function is $A(\boldsymbol{k},\omega)=\sum\limits_{s=\pm1} A^s(\boldsymbol{k},\omega)$ which follows from Eq.~(\ref{eqn:gf}) and is
\begin{equation}\label{eqn:akw}
A(\vect{k},\omega)=\sum\limits_{s=\pm} \frac{1}{\pi}\frac{-{\rm Im}\Sigma_s(\vect{k},\omega)}{[\omega -{\rm Re}\Sigma_s(\vect{k},\omega) -\epsilon^s_{\vect{\bar{k}}}]^2 + [{\rm Im}\Sigma_s(\vect{k},\omega)]^2}.
\end{equation}
Without self energy, the spectral density in Eq.~(\ref{eqn:akw}) reduces to a sum of two Dirac delta functions, and the strained dispersion curves are a simple geometrical scaling of the momentum axis applied to the bare unstrained dispersions.  When one has interactions, or finite self energies, as in Eq.~(\ref{eqn:akw}), one no longer obtains the strained spectral density by a simple geometric scaling of momentum of the unstrained spectral function.  
Of key importance in the discussion that follows is that Eq.~(\ref{eqn:akw}) gives results which differ from simply taking the spectral density in the unstrained case, which includes EEIs, and then scaling the momentum by an appropriate strain factor of $|\vect{B}|$.  Thus, we will use this simple geometrical scaling as a comparison to accentuate the impact of strain on correlation effects.

\section{Results}\label{sec:results}
\begin{figure}
  \begin{center}
  \includegraphics[width=0.98\linewidth]{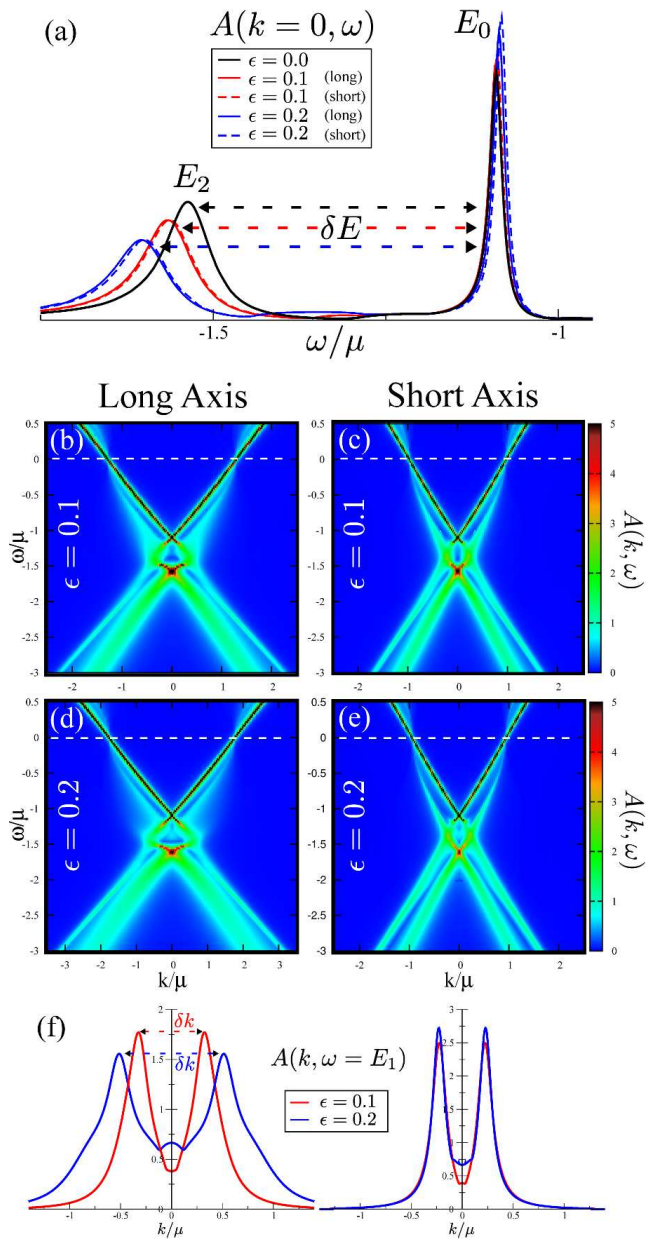}
  \end{center}
  \caption{\label{fig:cp}(Color online) (a)$A(k=0,\omega)$ for the long and short axis for increasing strain. $\delta E$ is robustly defined as the energy separation between the two dominant peaks.  The shift in $E_2$ would not occur in a non-interacting system.  (b$\rightarrow$e) Color plots representing $A(k,\omega)$ for cuts along the long and short axis for $\epsilon=0.1$ and 0.2.  (f) Complementary to the color plots:  A cut through the diamond structure at $\omega=E_1=(E_0+E_2)/2$.  The shift in $\delta k$ along the long axis is labelled.  }
\end{figure}
\begin{figure}
  \begin{center}
  \includegraphics[width=0.95\linewidth]{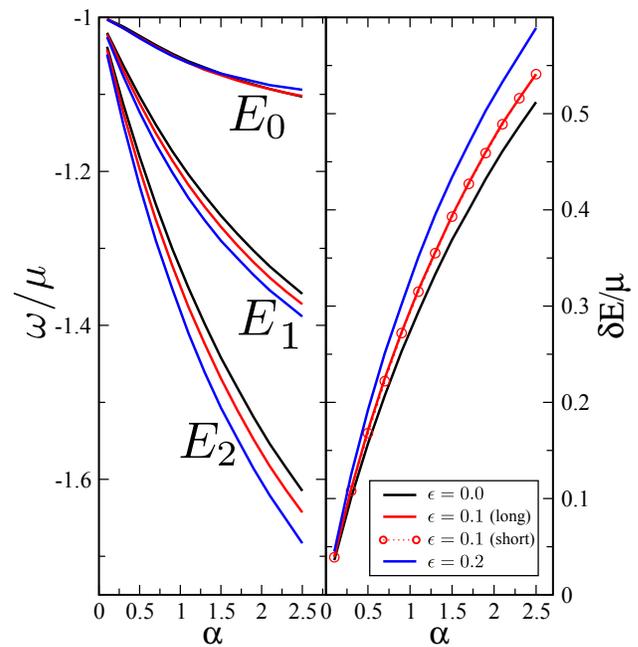}
  \end{center}
  \caption{\label{fig:delelong}(Color online) Left frame:  Peaks in $A(k=0,\omega)$ as a function of EEI coupling strength for strains of $\epsilon=0.0$, 0.1, and 0.2.  Peaks are identified as $E_0$, $E_1$ and $E_2$.  Right frame:  The energy width of the EEI diamond structure, defined as $\delta E=E_2-E_0$.  }
\end{figure} 
\begin{figure}
  \begin{center}
  \includegraphics[width=0.95\linewidth]{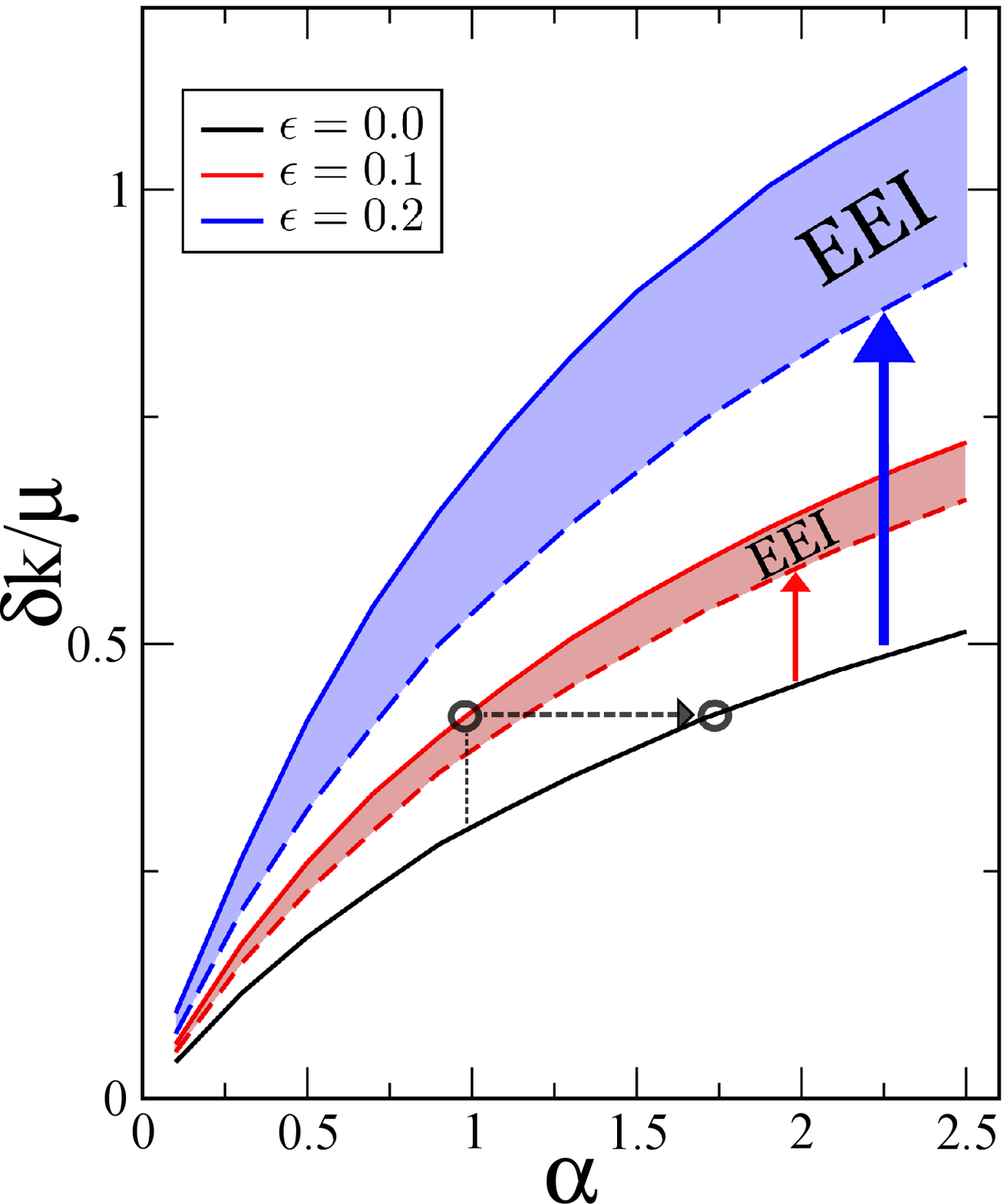}
  \end{center}
  \caption{\label{fig:delklong}(Color online) Extracted values of $\delta k$ along the long axis as a function of EEI coupling strength, $\alpha$, for unstrained $\epsilon=0.0$, and strained graphene, $\epsilon= 0.1$ and 0.2.  $\delta k$ is identified as in Fig.~\ref{fig:cp}(f).  Dashed lines are $\delta k$ measured in the simple geometric scaling case for scaling factors relevant to strain values of $\epsilon=0.1$ and 0.2 (red and blue dashed respectively).  The shaded regions accentuate the additional contributions to $\delta k$ from the correlation effects due to strain. }
\end{figure}  
Shown in Fig.~\ref{fig:cp} is an array of cuts along each of the long and short axis of the spectral function of a strained graphene sheet for the 10\% and 20\% strained cases for an EEI coupling strength of $\alpha=2$ (a value relevant for hydrogen passivated silicon carbide, H-SiC).\cite{bostwick:2010} 
We begin our discussion by considering the slice in energy through the $k=0$ momentum point.  In a non-interacting system, one would expect that the spectral function would exhibit no modification at the Dirac point from the application of strain.  However, with interactions, the application of strain modifies the local electronic structure which produces additional renormalizations not due to the simple geometry of the strain.  We show a specific interacting case in Fig.\ref{fig:cp}(a).  While there is no directional dependence at $k=0$ which would distinguish the long and the short axial directions, there is nevertheless a noticeable strain dependence.  Interestingly there is little to no modification of the primary Dirac point identified by $\omega=E_0$.  At larger negative energies the peak located at $E_2$ shifts significantly, resulting in a $\delta E$ which grows under the application of strain.

The full calculation along the long and short axis are shown in Fig.~\ref{fig:cp}(b-e).  One observes that the renormalizations do not influence the location of the Fermi level at $\omega=0$ (shown by the white dashed line) beyond the simple geometric modification along the momentum axis as discussed in Sec.~\ref{sec:selfe}.  At higher energies there are distinct features which go beyond a simple scaling of $\vect{k}$.  To help see this, we take a characteristic slice of the colour plots through the energy $E1$ which we define as the midpoint between $E_0$ and $E_2$.  We then have a robust way of identifying the radius of the plasmaron ring, $\delta k$, and how it disperses with strain, $\epsilon$, and the EEI coupling strength $\alpha$.
These slices are shown in Fig.~\ref{fig:cp}(f).
The application of strain results in an increase in $\delta k$ along the long axis (left frame) and little change for the short axis (right frame).  The later effect in the short axis is simply a manifestation of the relative values of $c_\parallel$ and $c_\perp$.  Along the long axis, one immediately sees features which do not allow the various strained cases to scale onto each other as would be the case without the EEI renormalizations.  As $\epsilon$ increases from 0.1 to 0.2, $\delta k$ increases and there is additional broadening of the peaks.  Also, there is a change in the value of the spectral function along $k=0$ which is important to note, since any modification at $k=0$ is significant  as it relates to changes in correlation effects.

Continuing the discussion of the $k=0$ variation, we focus on the dispersive nature of the two independent energy features, $E_0$ and $E_2$, which are identified as peaks in $A(k=0,\omega)$ as shown for an example in Fig.~\ref{fig:cp}(a).  In Fig.~\ref{fig:delelong} we show our results for these extracted peaks as a function of the EEI coupling strength, $\alpha$, as well as for variation in strain.  There is no difference between the long and short axis properties for this $k=0$ cut.  As noted, we find that strain shifts $E_2$ to larger negative energies while $E_0$ remains essentially unchanged.  This is interesting in that we can identify the upper cone which extends from $k=0$ at $\omega=E_0$ to $k=k_F$ at $\omega=0$ in which, even with the strain-corrected self energy, we see only the simple modification due to strain.  This is not so for other regions of $\vect{k}$ and $\omega$, near $E_2$.  Physically this represents how different regions\cite{wunsch:2006,hwang:2007} of the inverse dielectric function are sampled in Eqs.~(\ref{eqn:res}-\ref{eqn:line}) and suggests that $E_2$ is a feature dominated by coupling to the modified plasmaron band of Eq.~(\ref{eqn:pifix}) while $E_0$ (and the band extending towards $\vect{k_F}$) is primarily due to screened electron-electron coupling in the intra or interband regions\cite{wunsch:2006} and therefore is modified in a simple way by the application of strain.
The dispersive nature of $E_2$ results in a dependence of $\delta E$ on the strain amplitude, shown in the right hand frame of Fig.~\ref{fig:delelong}, which for small $\epsilon$ is expected to be a linear change from the unstrained case.\cite{pellegrino:2012}

Next we address the momentum scaling of the plasmaron ring  in more detail.  We plot in Fig.\ref{fig:delklong} how $\delta k$ is modified along the long axis with applied strain modulus, $\epsilon$, and with variation in EEI coupling strength, $\alpha$.  
We restrict $\alpha$ to a range of values where plasmaron bands have been observed experimentally\cite{walter:2011} and emphasize that larger values of $\alpha$ do not produce as clear features of the plasmaron band due to increased scattering as explored in previous calculations on unstrained graphene.\cite{principi:2012,dassarma:2013}
Shown in black for the unstrained case, this curve  agrees  precisely with previous calculations by Bostwick et a.\cite{bostwick:2010} which were found to agree well with experimental results.  Of primary interest in this work is the impact of strain, shown for $\epsilon=0.1 $ and 0.2 (red and blue solid lines respectively) across the same range of $\alpha$.  We compare these to the same feature in the case of a simple geometric scaling of momentum which introduces a $1/|\vect{B}|=1/c_\parallel$ scaling in $\delta k$ shown as the dashed curves.  We highlight the difference between the simple scaling result and the full G$_0$W-RPA calculation of the peak to peak distance $\delta k$.  The impact of the strained electron-electron interaction is to push the value of $\delta k$ to an even larger value than with momentum scaling alone.  Finally, we mark an example (open circles) where the application of 10\% strain to graphene at an EEI coupling strength of $\alpha=1$ produces a plasmaron ring with an effective $\delta k/\mu$ equivalent to an unstrained sheet of graphene with an $\alpha$ value nearly 70\% larger.

\section{Conclusions}\label{sec:conclusions}
We have studied the dependence of the electron-plasmon scattering structures of the single particle spectral density, for variation in EEI coupling strength with a uniaxially strained Dirac dispersion at finite chemical potential.  We have derived and calculated the G$_0$W-RPA self energy and full spectral density under strain and compared to the case of a simple geometric scaling of momenta as would be valid in a non-interacting system. 
We find that along the long axis, the strained EEI acts to modify the Dirac point separation energy, $\delta E$, and also to create an enhancement of the geometrical strain modification of the primary feature of electron-plasmon coupling, the radius of the `plasmaron ring', $\delta k$.  In addition, we find that these extra strain-induced renormalizations have very little effect on quasiparticle peaks close to the Fermi level.
By probing the regions in $\vect{k}$ and $\omega$ where strain has an effect on the correlations in the spectral density, it should be possible to isolate the changes in many-body effects from those due solely to geometry and sample, such as substrate effects which maybe be of concern in ARPES experiments.
Finally, we suggest that, in so far as the plasmaron ring is concerned, the application of strain could be used to tune the effective fine structure constant along the axis of applied strain in lieu of changing the sample substrate.  This concept will be very useful for any momentum sensitive probe where the signal in the direction of strain can be isolated, such as ARPES.  
  

\begin{acknowledgments}
This research was supported in part by the Natural Sciences and
Engineering Research Council of Canada (NSERC) and the Canadian Institute
for Advanced Research (CIFAR). 
\end{acknowledgments}


\bibliographystyle{apsrev4-1}
\bibliography{bib}

\end{document}